\documentclass[pra,twocolumn,superscriptaddress,showpacs,nofootinbibfloatfix,amsmath,amsfonts,amssymb]{revtex4-1}%
\usepackage{amsmath,amsfonts,amssymb,color}
\usepackage{amsthm}
\usepackage{leftidx}
\usepackage{graphicx}
\usepackage{xcolor}
\usepackage{dcolumn}
\usepackage{bm}
\usepackage{epstopdf}
\usepackage{epsfig}
\usepackage{environ}
\usepackage{pdfcomment}

\usepackage{float}
\usepackage[T1]{fontenc}
\usepackage[latin9]{inputenc}
\usepackage{setspace}
\usepackage{esint}

\begin{document}
	
\preprint{APS/123-QED}

\title{Non-Hermitian quasicrystal in dimerized lattices}

\author{Longwen Zhou}
\email{zhoulw13@u.nus.edu}
\affiliation{%
	College of Physics and Optoelectronic Engineering, Ocean University of China, Qingdao, China 266100
}
%\affiliation{%
%	Department of Physics, National University of Singapore, Singapore 117543
%}
\author{Wenqian Han}
\affiliation{%
	College of Physics and Optoelectronic Engineering, Ocean University of China, Qingdao, China 266100
}
%\author{Jiangbin Gong}%
%\email{phygj@nus.edu.sg}
%\affiliation{%
%	Department of Physics, National University of Singapore, Singapore 117543
%}

\date{\today}

\begin{abstract}
Non-Hermitian quasicrystals possess ${\cal PT}$ and metal-insulator
transitions induced by gain and loss or nonreciprocal effects. In
this work, we uncover the nature of localization
transitions in a generalized Aubry-Andr\'e-Harper model with dimerized
hopping amplitudes and complex onsite potential. By investigating the
spectrum, adjacent gap ratios and inverse participation ratios,
we find an extended phase, a localized phase and a mobility edge
phase, which are originated from
the interplay between hopping dimerizations and non-Hermitian onsite
potential. The lower and upper bounds of the mobility edge are further
characterized by a pair of topological winding numbers, which undergo
quantized jumps at the boundaries between different phases. Our discoveries thus
unveil the richness of topological and transport phenomena in dimerized
non-Hermitian quasicrystals.
\end{abstract}

%Use showkeys class option if keyword
\maketitle

\section{Introduction\label{sec:Intro}}
Quasicrystals have long-range order without spatial periodicity.
They form a class of system in between
crystals and fully disordered lattices~\cite{QCRev1,QCRev2,QCRev3,QCRev4,AAH1,AAH2}.
Experimentally, quasicrystals have been realized in a variety
of solid state materials and quantum simulators~\cite{QCExp1,QCExp2,QCExp3,QCExp4,QCExp5,QCExp6,QCExp7}.
Rich phenomena induced by quasiperiodicity have been revealed,
such as topological phases~\cite{TQC1,TQC2,TQC3,TQC4,TQC5,TQC6,TQC7,ZhouCDHM},
quantized adiabatic pumping~\cite{TQC2,TQC7,ZhouCDHM}, anomalous
transport and localization transitions~\cite{MIT2,MIT3,MIT4,MIT5,MIT6,MIT7,MIT8,MIT9,MIT10,MIT11,MIT12,MIT13,MIT14,MIT15,MIT16,MIT17,MIT18,MIT19,MIT20,MIT21,MIT22,MIT23,MIT24,MIT25,MIT26,MIT27,MIT28,MIT29,MIT30,MIT31},
attracting attention over a broad range of research fields.

Recently, the study of quasicrystals has been extended to
non-Hermitian systems, where the interplay between quasiperiodicity
and gain/loss or nonreciprocity could induce exotic dynamical, localization
and topological phenomena~\cite{LonghiQC1,LonghiQC2,LonghiQC3,LonghiQC4,ChenQC1,ChenQC2,ChenQC3,ChenQC4,ChenQC5,ChenQC6,CaiQC1,NHQC1,NHQC2,NHQC3,NHQC4,NHQC5,NHQC6,NHQC7}.
In particular, in non-Hermitian variants of the Aubry-Andr\'e-Harper
(AAH) model, complex onsite potential or nonreciprocal
hopping could induce a ${\cal PT}$-breaking transition and a metal-insulator transition, which can be further characterized
by a spectral topological winding number~\cite{LonghiQC1,ChenQC1}. In related studies, non-Hermiticity
induced mobility edges in generalized and superconducting AAH models
have also been found and described by topological invariants~\cite{LonghiQC3,ChenQC2,ChenQC3,ChenQC4,CaiQC1,NHQC5}.
Besides, the investigation of wavepacket spreading in non-Hermitian
quasicrystals~(NHQCs) have revealed their anomalous dynamical
features~\cite{LonghiQC4,ChenQC5}, such as the disordered-enhanced transport~\cite{LonghiQC4}.

Meanwhile, the Su-Schrieffer-Heeger (SSH) model~\cite{SSH1}
provides another paradigm for the study of localization~\cite{SSHAL1},
topological~\cite{SSHTP1} and non-Hermitian~\cite{NHTPRev1,ZhouNH1,ZhouNH2}
physics. Specially, the hopping dimerization
allows the SSH model to possess a topological phase characterized
by an integer winding number and degenerate edge
modes~\cite{SSHTP1}. However, in the context of non-Hermitian
quasicrystals, phases and phenomena that could arise
due to the interplay between spatial aperiodicity and hopping dimerization
have not been revealed. In this manuscript, we address this issue
by introducing a dimerized quasiperiodic lattice in Sec.~\ref{sec:Mod},
which forms a hybridization of the SSH and non-Hermitian
AAH models. The system is found to possess rich patterns of
${\cal PT}$-breaking and localization transitions, together with three
phases of distinct transport nature. In Sec.~\ref{sec:Res},
we perform detailed analyses of the spectrum and localization nature
of these phases, and construct a pair of topological winding numbers
to characterize the transitions between them. Despite
an extended and a localized phase, we also find a mobility edge phase, 
which is absent without the hopping dimerization. 
These discoveries thus uncover the richness of topological and transport
phenomena in dimerized NHQCs. In Sec.~\ref{sec:Sum},
we summarize our results and discuss potential future directions.
Further details about the spectrum, Lyapunov exponents and wavepacket dynamics
are provided in the Appendixes \ref{subsec:Edetail}--\ref{subsec:Dyn}.

\section{Model\label{sec:Mod}}

In this section, we introduce the dimerized NHQC that will be
investigated in this work. Our model can be viewed as an extension
of the AAH model \cite{AAH1,AAH2}, which is prototypical
in the study of localization transitions in one-dimensional (1D) quasicrystals.
In position representation, the Hamiltonian of the AAH model is
$\hat{H}_{{\rm AAH}}=\sum_{n}\left(J|n\rangle\langle n+1|+{\rm H.c.}+V\cos(2\pi\alpha n+\lambda)|n\rangle\langle n|\right)$,
where $\{|n\rangle\}$ represents the eigenbasis of the lattice, $J$
is the hopping amplitude, $V$ controls the strength
of onsite potential, and $\lambda$ is a phase
shift. When $\alpha$ is irrational, the
potential $V_{n}=V\cos(2\pi\alpha n+\lambda)$ is quasiperiodic in
$n$, and $\hat{H}_{{\rm AAH}}$ describes a 1D quasicrystal. When $V<2J$, the
spectrum of $\hat{H}_{{\rm AAH}}$ is continuous and all its eigenstates
are extended under the periodic boundary condition (PBC). Comparatively,
$\hat{H}_{{\rm AAH}}$ possesses a point spectrum with localized eigenstates
at all energies when $V>2J$. When $V=2J$, the spectrum of $\hat{H}_{{\rm AAH}}$
is purely singular continuous with critical wavefunctions, and the
system undergoes a localization transition when
$V$ changes from $V<2J$ to $V>2J$~\cite{QCRev4}.

Recently, a non-Hermitian variant of $\hat{H}_{{\rm AAH}}$ was introduced
by setting $\lambda=\beta+i\gamma$, with $(\beta,\gamma)\in\mathbb{R}$~\cite{LonghiQC1}.
Such a non-Hermitian AAH model possesses ${\cal PT}$-breaking and
localization transitions at $\gamma_{c}=\ln(2J/V)$, which are
accompanied by the quantized jump of a spectral winding number
$w$. For $\gamma<\gamma_{c}$ ($\gamma>\gamma_{c}$), the spectrum is real (complex) 
with winding number $w=0$ ($w=-1$), and each eigenstate is extended (localized)~\cite{LonghiQC1}. 
Since all states
subject to the same localization transition at $\gamma=\gamma_{c}$,
no mobility edges are found. Similar results are
reported in a nonreciprocal AAH model~\cite{ChenQC1}, which might be
related to the model in Ref.~\cite{LonghiQC1} by Fourier transformations.

\begin{figure}
	\begin{centering}
		\includegraphics[scale=0.38]{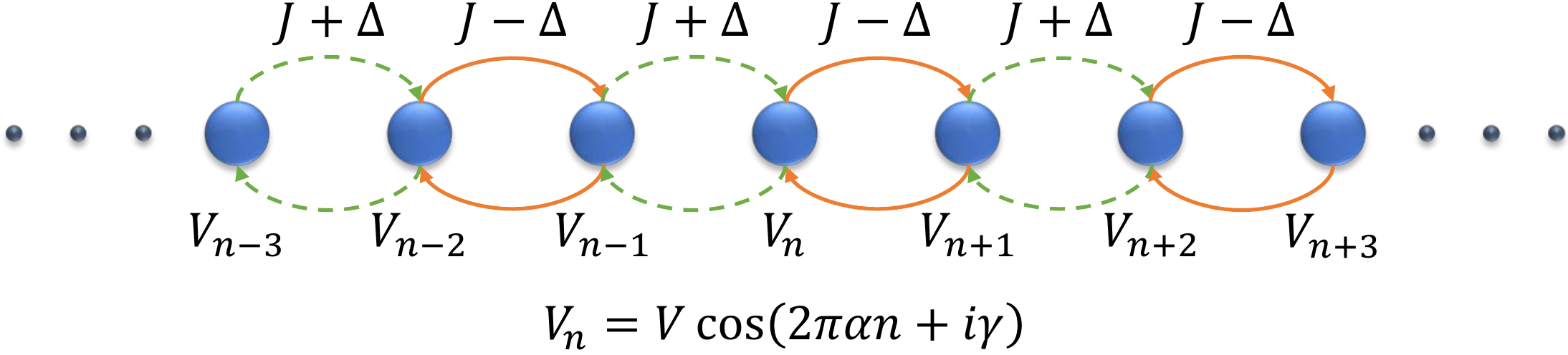}
		\par\end{centering}
	\caption{Schematic illustration of the dimerized NHQC. Solid
		balls denote lattice sites with index $n\in\mathbb{Z}$. 
		Hopping amplitudes are alternated between $J-\Delta$ (solid
		lines) and $J+\Delta$ (dashed lines). $V_{n}$
		represents the strength of onsite potential, with irrational
		modulation parameter $\alpha$ and phase shift $i\gamma$.\label{fig:Sketch}}
\end{figure}

In this work, we consider another extension of the AAH quasicrystal
by introducing hopping dimerizations. We set
$J\rightarrow J_{n}=J+(-1)^{n}\Delta$ as the hopping amplitude between
the $n$'s and the $(n+1)$'s lattice sites, and let $\lambda=i\gamma$
be the imaginary phase shift in the onsite potential $V_{n}$.
Our model Hamiltonian thus takes the form
\begin{alignat}{1}
\hat{H} & =\sum_{n}\left\{ \left[J+(-1)^{n}\Delta\right]|n\rangle\langle n+1|+{\rm H.c.}\right\} \nonumber \\
& +\sum_{n}V\cos(2\pi\alpha n+i\gamma)|n\rangle\langle n|.\label{eq:H}
\end{alignat}
A schematic illustration of the model is shown in Fig.~\ref{fig:Sketch}.
When $V=0$, Eq.~(\ref{eq:H}) reduces to the SSH model with hopping
dimerization $\Delta$. With $V,\gamma\neq0$,
$\hat{H}$ describes an SSH chain with quasiperiodically
correlated non-Hermitian disorder. Therefore, our system realizes a
hybridization between the AAH and SSH models. 
Since the realizations of SSH and non-Hermitian
AAH models have both been discussed~\cite{LonghiQC1,ExpSetup2}, we expect
our model to be within reach under current or near term experimental
conditions. Furthermore, our construction allows us to explore the
interplay between hopping dimerizations and quasiperiodic non-Hermitian
potential, which could lead to rich phase diagrams and transport phenomena, 
as will be shown in the following section.

\section{Results}\label{sec:Res}

In this section, we investigate the
dimerized NHQC in Eq.~(\ref{eq:H}) from the perspective of
spectrum, states, and topological invariants. In Sec.~\ref{subsec:E},
we study the ${\cal PT}$-transitions in our system 
with respect to the hopping
dimerization $\Delta$ and imaginary phase shift $i\gamma$. 
Adjacent gap ratios (AGRs) in the real part of the spectrum are found to exhibit three
distinct regions of level statistics, 
implying the existence of an extended, a localized and a mobility edge
phase in which extended and localized states coexist. In Sec.~\ref{subsec:IPR},
we study the inverse participation ratios~(IPRs) of the states in our system, and confirm the presence
of a mobility edge region between the extended and localized phases in
the parameter space, which is originated from the interplay between
the dimerized hopping and complex onsite potential. Using the critical energies of
the mobility edge, we construct a pair of topological winding numbers
in Sec.~\ref{subsec:WN}, which could fully characterize the transitions
between different phases. Throughout
this section, we set the onsite modulation $\alpha=\frac{\sqrt{5}-1}{2}$
as the inverse of golden ratio to realize the quasiperiodicity
of the potential. Under the PBC, we
take the rational approximation of $\alpha$ by setting $\alpha\simeq p/q$,
with $p,q$ being two adjacent terms ($p<q$) in the Fibonacci
sequence, and set the size of lattice $L=q$. We also let $J=1$
be the unit of energy and set the Planck constant $\hbar=1$.
All system parameters are given in dimensionless units.

\subsection{Spectrum and level statistics}\label{subsec:E}

The spectrum of the dimerized NHQC is obtained by solving the
eigenvalue equation $\hat{H}|\psi\rangle=E|\psi\rangle$. Projecting
the equation to the lattice representation, we find
\begin{equation}
\left[J+(-1)^{n}\Delta\right]\psi_{n+1}+\left[J+(-1)^{n-1}\Delta\right]\psi_{n-1}+V_{n}\psi_{n}=E\psi_{n},\label{eq:Seq}
\end{equation}
where $V_{n}=V\cos(2\pi\alpha n+i\gamma)$. Since $V_{n}=V_{-n}^{*}$,
$\hat{H}$ possesses the ${\cal PT}$-symmetry and its spectrum can be real
in the ${\cal PT}$-invariant region. With the increase of $\gamma$, one would
expect a transition through which the ${\cal PT}$-symmetry is broken and the
spectrum changes from real to complex. 
If $|\psi\rangle$ is an eigenstate
of $\hat{H}$ with energy $E$, ${\cal PT}|\psi\rangle$ should
be an eigenstate with energy $E^{*}$. Therefore, the spectrum
of $\hat{H}$ is symmetric with respect to the real axis
on the complex $E$ plane. Furthermore, applying the chiral~(sublattice)
symmetry operator $\Gamma=\sum_{n}(-1)^{n-1}|n\rangle\langle n|$
to the eigenvalue equation, we obtain $\hat{H}'\Gamma|\psi\rangle=(-E)\Gamma|\psi\rangle$,
where
$\hat{H}'=\sum_{n}(\left[J+(-1)^{n}\Delta\right]|n\rangle\langle n+1|+{\rm H.c.}+V\cos(2\pi\alpha n+\pi+i\gamma)|n\rangle\langle n|)$.
Since $\alpha$ is irrational, $2\pi\alpha n\,\mod\,2\pi$ uniformly
fills the range of $[0,2\pi)$ for $n=1,...,L$ in
the thermodynamic limit $L\rightarrow\infty$. Therefore, it is possible
to remove the extra phase shift $\pi$ in $\hat{H}'$ by resetting
the origin of the lattice. $\hat{H}$ and $\hat{H}'$ thus share the
same spectrum in the limit $L\rightarrow\infty$. This
implies that both $E$ and $-E$ are eigenvalues of $\hat{H}$,
and the spectrum of $\hat{H}$ is symmetric with respect to the
imaginary axis on the complex $E$ plane. These generic
features are confirmed by numerical results reported
in Appendix \ref{subsec:Edetail}.

\begin{figure}
	\begin{centering}
		\includegraphics[scale=0.45]{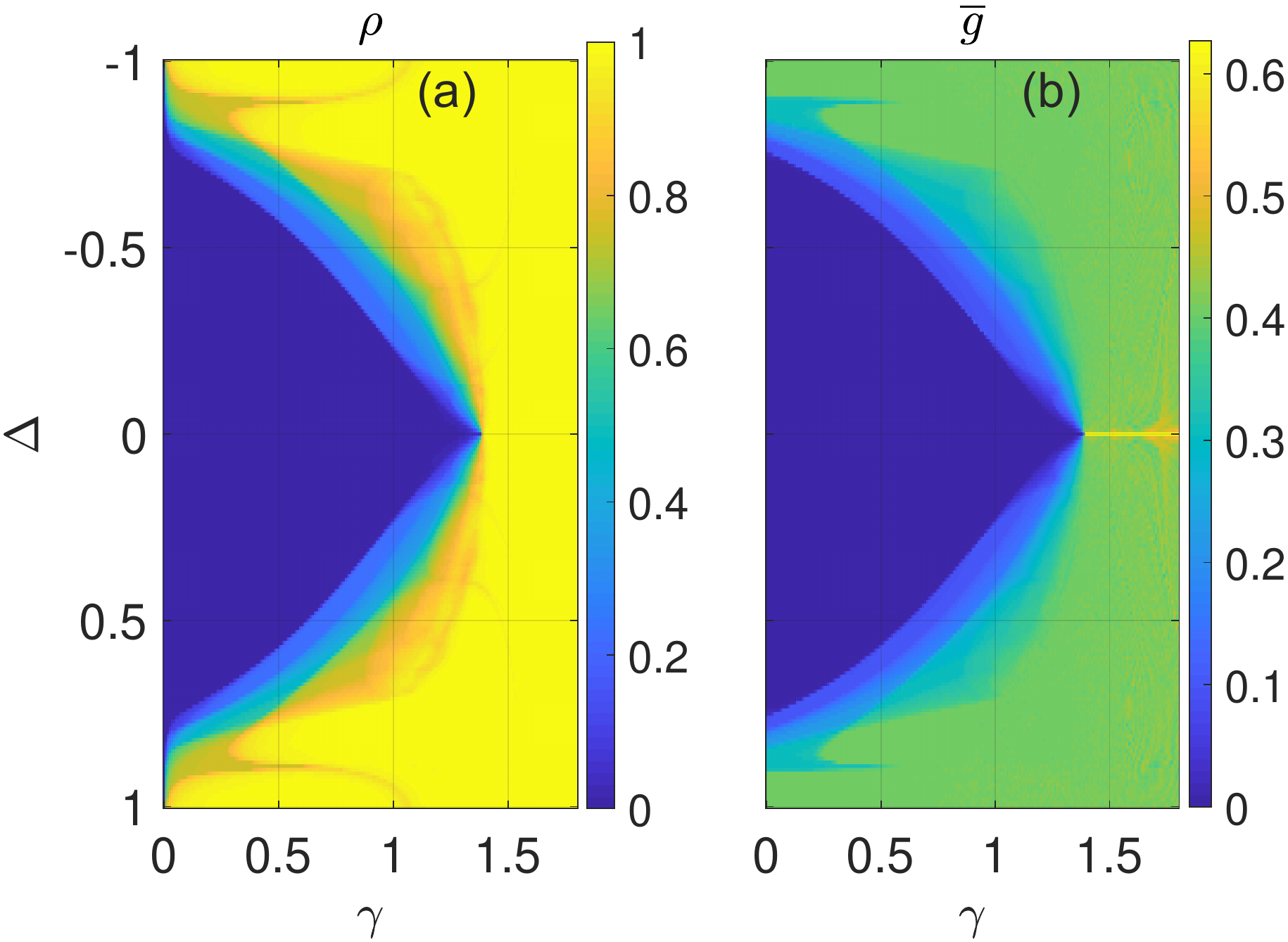}
		\par\end{centering}
	\caption{Density of states with complex eigenvalues and averaged AGRs
		of $\hat{H}$ versus the dimerization amplitude $\Delta$ and
		the imaginary part of phase shift $\gamma$ in (a) and (b),
		respectively. System parameters are set as $J=1$, $V=0.5$,
		and $\alpha=\frac{\sqrt{5}-1}{2}$. PBC is taken in the calculation
		and the length of lattice is $L=2584$.\label{fig:DOSAGR}}
\end{figure}

To have a comprehensive view of the spectrum and its connection
with localization properties of states, we
present the density of states with nonzero imaginary parts of energies 
and the mean of AGRs
versus $\gamma$ and $\Delta$ in Figs.~\ref{fig:DOSAGR}(a) and \ref{fig:DOSAGR}(b),
respectively. The density of states with complex energies is
defined as
\begin{equation}
\rho\equiv\frac{1}{L}\sum_{j=1}^{L}N(|{\rm Im}E_{j}|).\label{eq:DOS}
\end{equation}
Here $L$ is the length of lattice. For the $j$th eigenstate of $\hat{H}$
with energy $E_{j}$, we have $N(|{\rm Im}E_{j}|)=1$ ($=0$)
if $|{\rm Im}E_{j}|>0$ ($|{\rm Im}E_{j}|=0$). The AGR
is defined as
$g_{j}\equiv\frac{\min(\epsilon_{j},\epsilon_{j+1})}{\max(\epsilon_{j},\epsilon_{j+1})}$,
where $\epsilon_{j}={\rm Re}E_{j}-{\rm Re}E_{j-1}$ denotes the spacing
between the real parts of energies of the $(j-1)$'s and the $j$'s
eigenstates of $\hat{H}$ in ascending order. According to the non-Hermitian random-matrix theory
\cite{NHRMT1,NHRMT2,NHRMT3,NHRMT4,NHRMT5}, the statistical property
of $g_{j}$ is closely related to the localization features of non-Hermitian
disordered systems. We consider
the mean value of AGRs over all eigenstates \cite{ChenQC2},
which is defined as
\begin{equation}
\overline{g}=\frac{1}{L}\sum_{j}g_{j}.\label{eq:MAGR}
\end{equation}
The values of $\overline{g}$ in different parameter regions could
provide us with a guideline to distinguish phases with distinct localization
nature in the dimerized NHQC.

In Fig.~\ref{fig:DOSAGR}(a), we find that all the eigenvalues of $\hat{H}$
are real in a finite region~(in dark blue) of the~$(\gamma,\Delta)$
parameter space, in which $\rho=0$ and the system is ${\cal PT}$-invariant.
With the increase of $\gamma$ and 
$\Delta$, the spectrum changes from real to complex, with
$0<\rho<1$, and the system enters a ${\cal PT}$-breaking phase. When
$\gamma$ is large enough, almost all states have complex eigenvalues
and we have $\rho\simeq1$. Notably, the boundary between the ${\cal PT}$-invariant
and ${\cal PT}$-broken phases of the spectrum in Fig.~\ref{fig:DOSAGR}(a)
is consistent with the boundary separating states with averaged AGRs
$\overline{g}\simeq0$ and $\overline{g}>0$ in Fig.~\ref{fig:DOSAGR}(b).
According to the general results of level statistics, all eigenstates
of $\hat{H}$ would be extended when $\overline{g}\simeq0$. Therefore,
we expect an extended phase in the ${\cal PT}$-invariant region of $\hat{H}$. 
Moreover, the region
with $\overline{g}>0$ in Fig.~\ref{fig:DOSAGR}(b) can be separated
into two distinct zones. When $\gamma$ is large enough, we find $\overline{g}\simeq0.4$
for all $\gamma$ with $\Delta\neq0$, which corresponds to a localized
phase according to the level statistics~\cite{NHRMT1,NHRMT2,NHRMT3,NHRMT4,NHRMT5}. 
In the region with $0<\overline{g}<0.4$,
$\overline{g}$ grows smoothly with the increase of $\gamma$
and $\Delta$, implying the existence of a mobility edge phase.
The emergence of such a phase is solely due to the hopping dimerization, which is absent in the original AAH model.
These observations are further confirmed
by the study of IPRs presented in the next subsection.

\subsection{Inverse participation ratio\label{subsec:IPR}}

To further unveil the localization nature of states in the dimerized
NHQC, we study the IPR of our system in this subsection.
For a normalized eigenstate $|\psi^{(j)}\rangle=\sum_{n}\psi_{n}^{(j)}|n\rangle$
of $\hat{H}$,
the IPR is defined in the lattice representation as
${\rm IPR}^{(j)}\equiv\sum_{n=1}^{L}|\psi_{n}^{(j)}|^{4}$,
where $L$ is the length of lattice. For 1D systems in the thermodynamic
limit, we generally have ${\rm IPR}\sim L^{-1}$ if $|\psi\rangle$
is an extended state, and ${\rm IPR}\sim\xi^{-1}$ if $|\psi\rangle$
is localized, where the localization length $\xi$ is independent
of $L$~\cite{CMTBook1}. Therefore, we expect that when $L$ is large,
all the eigenstates of $\hat{H}$ have ${\rm IPR}\simeq0$ in the
extended phase. To capture the localization transitions between
different phases, we further introduce the maximum and minimum of IPRs, which are defined as
\begin{alignat}{1}
\max({\rm IPR}) & \equiv\max_{j\in\{1,...,L\}}{\rm IPR}^{(j)},\label{eq:IPRmax}\\
\min({\rm IPR}) & \equiv\min_{j\in\{1,...,L\}}{\rm IPR}^{(j)}.\label{eq:IPRmin}
\end{alignat}
When $\max({\rm IPR})$ diverges from zero but $\min({\rm IPR})$
remains at zero, localized states start to appear and
the system switches from the extended to the mobility edge phase.
When $\min({\rm IPR})$
also deviates from zero, the last extended state vanishes and the
system enters a phase in which all states are localized.
The maximal and minimal values of IPR can thus be used to distinguish
phases with different localization nature and locate the corresponding
phase boundaries.

\begin{figure}
	\begin{centering}
		\includegraphics[scale=0.45]{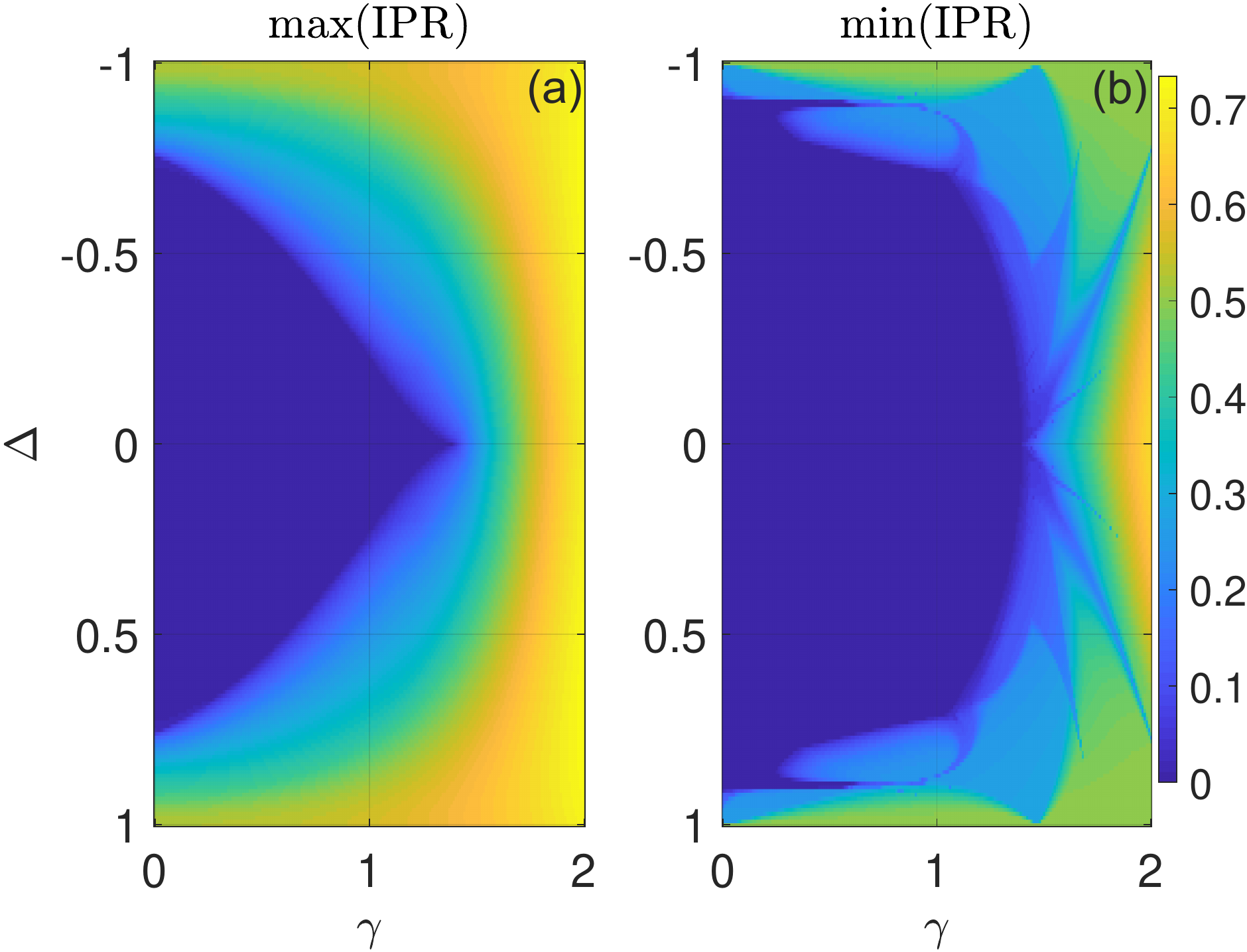}
		\par\end{centering}
	\caption{The maximum and minimum of IPRs versus the imaginary part of phase
		shift $\gamma$ and hopping dimerization $\Delta$ in (a) and
		(b). System parameters are set as $J=1$, $V=0.5$,
		and $\alpha=\frac{\sqrt{5}-1}{2}$. The length of lattice is $L=610$.
		PBC is taken in the diagonalization $\hat{H}$.\label{fig:IPR}}
\end{figure}

In Fig.~\ref{fig:IPR}, we present the maximum and minimum of IPRs
versus the imaginary phase shift and hopping dimerization. 
In Fig.~\ref{fig:IPR}(a), we find a region with $\max({\rm IPR})\simeq0$
(in dark blue) in the $\gamma$-$\Delta$ parameter space, which means
that all states of $\hat{H}$ for a given set of system parameters
in this region are extended. Notably, the scope
of this region is coincide with the regions in Fig.~\ref{fig:DOSAGR},
in which the density of states 
$\rho$ and averaged AGRs $\overline{g}$ vanish. Therefore,
the domain with $\max({\rm IPR})\simeq0$ 
indeed corresponds to a ${\cal PT}$-invariant extended phase, in which all eigenstates
of $\hat{H}$ have real energies and delocalized profiles.
In Fig.~\ref{fig:IPR}(b), we also observe a zone
with $\min({\rm IPR})\simeq0$ (in dark blue), in which extended
eigenstates of $\hat{H}$ persist up to its boundary. 
As expected, the zone with $\min({\rm IPR})\simeq0$
covers the region with $\max({\rm IPR})\simeq0$.
Beyond that, we have an intermediate region with $\min({\rm IPR})\simeq0$
and $\max({\rm IPR})>0$, before the system enters a localized
phase with $\min({\rm IPR})>0$. In the
intermediate region, since there are states with both vanishing and
finite IPRs, extended and localized states must coexist. 
The range of
this intermediate region is also consistent 
with the regions in Fig.~\ref{fig:DOSAGR} with $0<\rho<1$ and $0<\overline{g}<0.4$.
These observations confirm that there indeed exists a mobility edge
phase between the extended and localized phases of the dimerized NHQC, 
in which the ${\cal PT}$-symmetry is broken and yet only part of the states
are localized with complex energies. Note that this mobility edge
phase does not exist in the non-Hermitian AAH model with $\Delta=0$,
which highlights the importance of hopping dimerizations in creating
unique states of matter and localization transitions in non-Hermitian
quasicrystals.

To summarize, we find that the IPR can be viewed
as an ``order parameter'' to describe the localization transitions
in the dimerized NHQC. The behaviors of IPRs in distinct parameter
domains could be employed to characterize the three different
phases in the system. 
For completeness, we study the Lyapunov exponents of the system in Appendix \ref{subsec:LE}, and 
find consistent results as predicted by IPRs.
We have also checked the spectrum and IPRs of the system
under the open boundary condition~(OBC), and find consistent results with those
obtained under the PBC, excluding possible impact of non-Hermitian skin effects.
In the following, we demonstrate
that the localization transitions in our system are of topological
origin, and can be depicted by a pair of spectral winding numbers.

\subsection{Topological winding number\label{subsec:WN}}
In previous studies, spectral winding numbers have been employed to
characterize non-Hermitian topological matter in 1D clean and disordered
systems \cite{LonghiQC1,NHTPRev1}. For the dimerized NHQC, we can
introduce a pair of winding numbers to describe the topological
nature of its localization transitions. These numbers can
be defined as
\begin{equation}
w_{\ell}=\lim_{N\rightarrow\infty}\frac{1}{2\pi i}\int_{0}^{2\pi}d\beta\partial_{\beta}\ln\left\{ \det\left[H(\beta/N)-{\cal E}_{\ell}\right]\right\}.\label{eq:WN}
\end{equation}
$N$ is the number of dimerized cells of the lattice and $\ell=1,2$.
The phase shift $\beta/N$ is introduced into the
Hamiltonian $\hat{H}$ via setting $V_{n}\rightarrow V\cos(2\pi\alpha n+\beta/N+i\gamma)$.
$({\cal E}_{1},{\cal E}_{2})$ are two real-valued base energies,
and $(w_{1},w_{2})$ count the number of times the spectrum of $H(\beta/N)$
winds around these energies when $\beta$ sweeps over a cycle
from zero to $2\pi$. It is clear that $w_{1}$ ($w_{2}$) can
be nonzero only if the spectrum of $\hat{H}$ around ${\cal E}_{1}$
(${\cal E}_{2}$) take complex values. These winding numbers are thus
closely related to the complex spectrum structure of the system. When
the spectral does not possess a mobility edge, there is only a single
base energy that can in principle be chosen arbitrarily \cite{LonghiQC1},
and we would always have $w_{1}=w_{2}$. In our model this is the
case when $\Delta=0$ (uniform hopping). If mobility edges exist in
the spectral, the choice of base energies $({\cal E}_{1},{\cal E}_{2})$
should be related to its boundaries~\cite{LonghiQC3}.
More precisely, in a given range of imaginary phase shift $i\gamma$
or hopping dimerization $\Delta$, we choose ${\cal E}_{1}$ to be
the real part of energy of the first eigenstate of $\hat{H}$
whose IPR starts to deviate from zero, i.e., the first eigenstate
that becomes localized. Similarly, we set ${\cal E}_{2}$ as the real
part of energy of the last eigenstate whose profile
changes from extended to localized. ${\cal E}_{1}$ and ${\cal E}_{2}$
thus decide the lower and upper bounds of the mobility edge on the
${\rm Re}E$-$\gamma$ or ${\rm Re}E$-$\Delta$ plane. The winding
numbers $w_{1}$ and $w_{2}$ defined with respect to
${\cal E}_{1}$ and ${\cal E}_{2}$ are expected to have quantized
jumps when the mobility edge appears and vanishes in the spectrum
\cite{LonghiQC3}.

\begin{figure}
	\begin{centering}
		\includegraphics[scale=0.48]{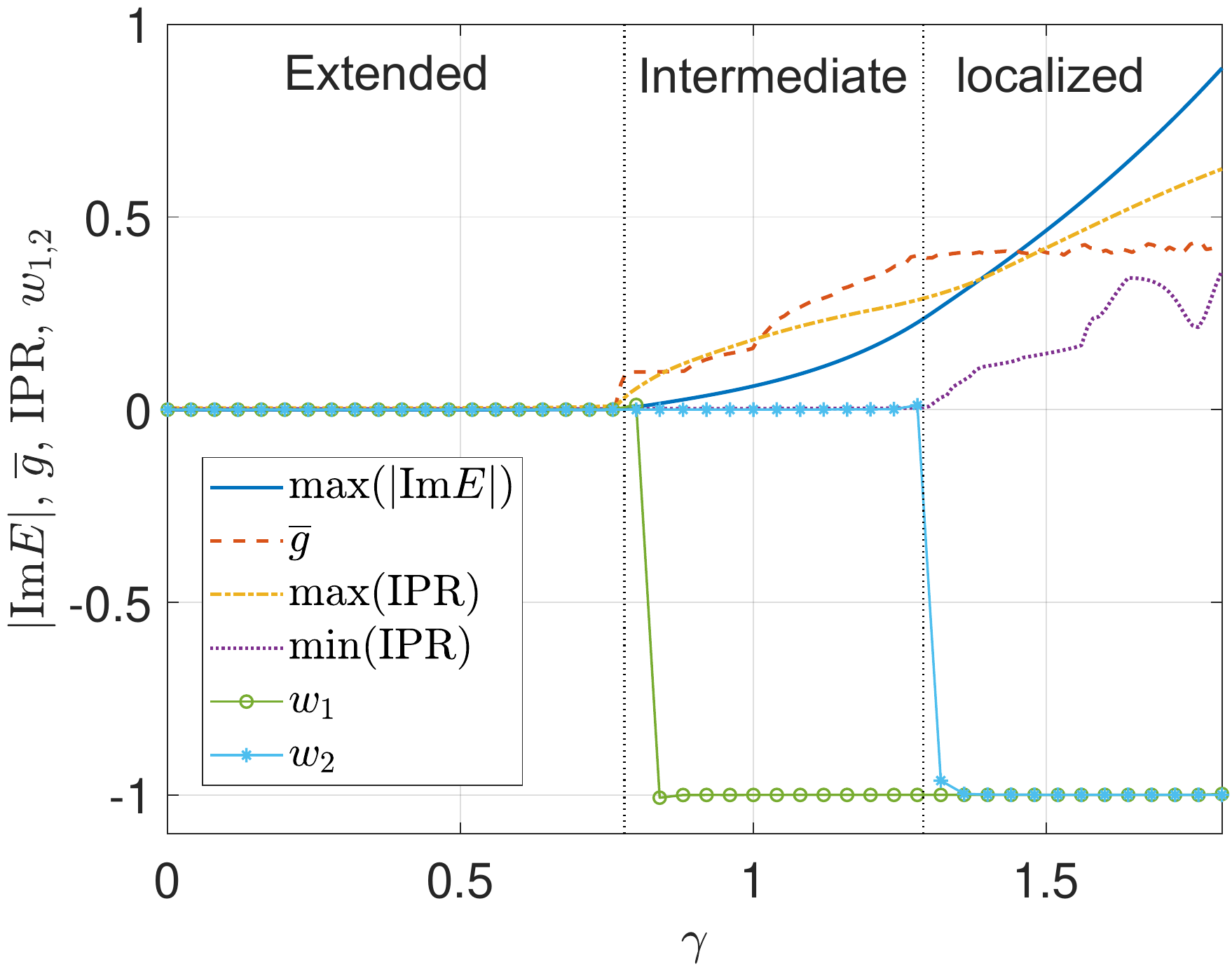}
		\par\end{centering}
	\caption{The winding numbers $(w_{1},w_{2})$ versus the imaginary part of
		phase shift $\gamma$. System parameters are set as $J=1$,
		$V=0.5$, $\Delta=0.4$, $\alpha=\frac{\sqrt{5}-1}{2}$. The size
		of lattice is $L=610$ with PBC. The vertical dotted lines highlight
		the critical values $\gamma_{c1}$ and $\gamma_{c2}$,
		where transitions between different phases happen.\label{fig:WN}}
\end{figure}

In Fig.~\ref{fig:WN}, we report the winding numbers,
the maximal imaginary parts of energy, the averaged AGRs
and the IPRs together for a typical
set of system parameters. We find that the invariants $(w_{1},w_{2})=(0,0)$
in the ${\cal PT}$-invariant extended phase,
with $\max(|{\rm Im}E|)\simeq\overline{g}\simeq\max({\rm IPR})\simeq\min({\rm IPR})=0$.
When the imaginary part of phase shift passes the first critical point
$\gamma_{c1}\approx0.78$, $w_{1}$ undergoes a
quantized jump from $0$ to $-1$, and $[\overline{g},\max({\rm IPR})]$
start to deviate from zero. The system thus enters a mobility edge
phase with winding numbers $(w_{1},w_{2})=(-1,0)$. Note that such
a phase is absent if $\Delta=0$,
which reveals the role of hopping dimerization in the creating of
new phases and phase transitions in the dimerized NHQC. When $\gamma$
goes through the second critical point $\gamma_{c2}\approx1.29$,
the winding number $w_{1}$ stays at $-1$ while $w_{2}$ jumps from
zero to $-1$. Meanwhile, $\min({\rm IPR})$ starts to deviate
from zero and the averaged AGR converges to $\overline{g}\simeq0.4$.
The system therefore gets into a phase with 
localized eigenstates at all possible energies and winding numbers $(w_{1},w_{2})=(-1,-1)$.
The invariants $(w_{1},w_{2})$ can thus
be used to distinguish the three possible phases with different localization
nature in the system, and characterize the transitions between them.

\begin{figure}
	\begin{centering}
		\includegraphics[scale=0.47]{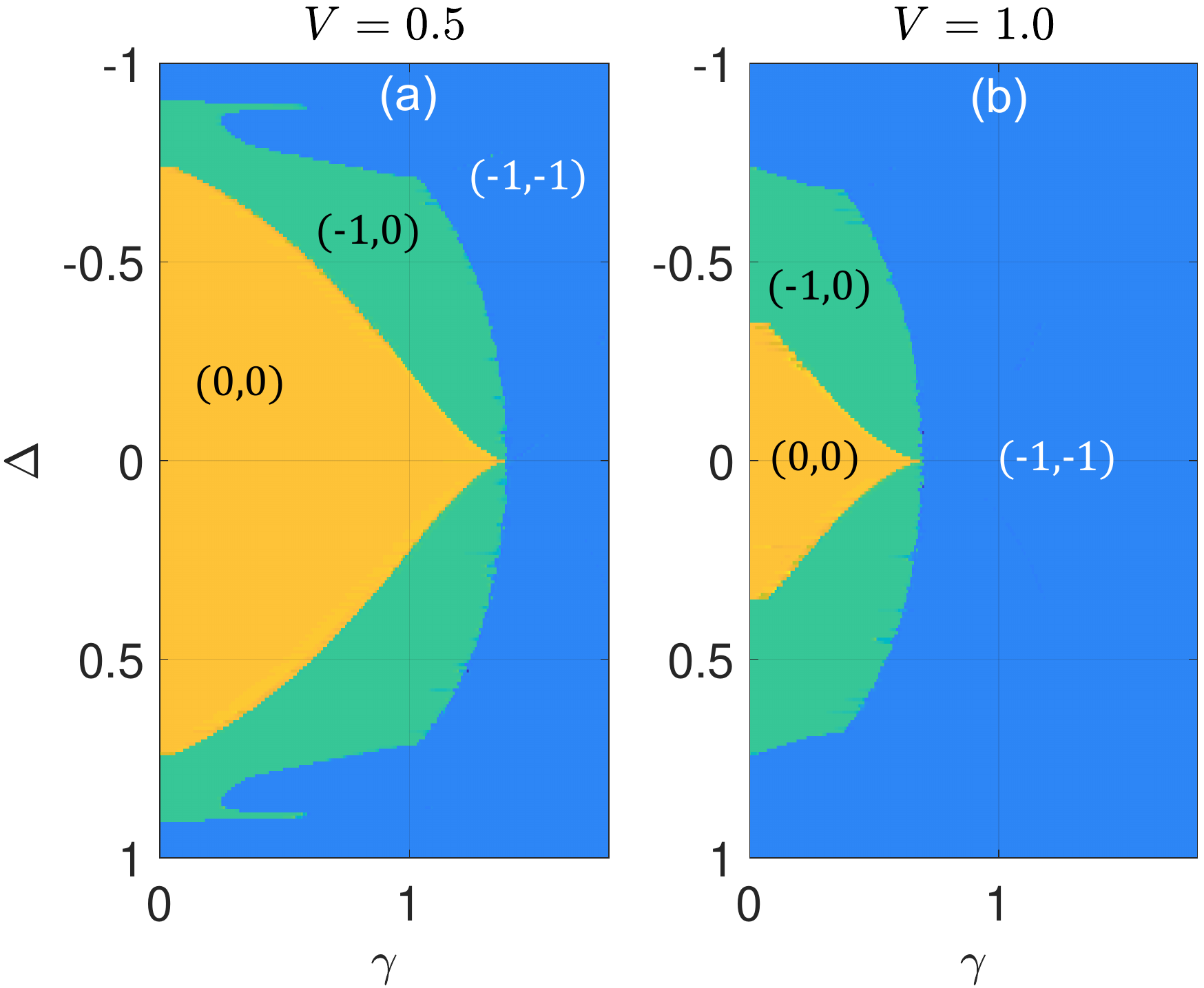}
		\par\end{centering}
	\caption{Topological phase diagrams of the dimerized NHQC. System parameters
		are set as $J=1$, $\alpha=\frac{\sqrt{5}-1}{2}$, and $V=0.5$
		($V=1$) for (a) {[}(b){]}. The length of lattice is
		$L=610$, containing $N=305$ dimerized cells. In (a)
		and (b), each region with a uniform color corresponds to a topological
		phase, whose winding numbers are denoted 
		therein. The extended, mobility edge and localized phases have
		$(w_{1},w_{2})=(0,0)$, $(-1,0)$ and $(-1,-1)$.
		$w_{1}$ or $w_{2}$ jumps at a boundary
		between different phases. PBC is taken in the calculation.\label{fig:PhsDiag}}
\end{figure}

In Fig.~\ref{fig:PhsDiag}, we present the topological phase diagrams
of the dimerized NHQC versus the imaginary part of phase shift
and hopping dimerization for two typical
sets of system parameters. In the region of $\rho\simeq0$ {[}real
spectrum, see also Fig.~\ref{fig:DOSAGR}(a){]} and $\max({\rm IPR})\simeq0$
{[}all bulk states are extended, see also Fig.~\ref{fig:IPR}(a){]},
we find $w_{1}=w_{2}=0$, implying
that the ${\cal PT}$-invariant extended phase is topologically trivial according
to the spectral winding numbers. When 
$\rho$ and $\max({\rm IPR})$ start to deviate from zero, the system
undergoes a ${\cal PT}$-breaking transition and enters a mobility edge phase.
In the meantime, $w_{1}$ takes a quantized jump from $0$ to
$-1$, whereas $w_{2}$ remains at zero in Fig.~\ref{fig:PhsDiag}(a).
The transition from extended to mobility edge phases is
thus topological and captured by the quantized change of winding number
$w_{1}$. The mobility edge phase can also be viewed as a topological
phase characterized by $(w_{1},w_{2})=(-1,0)$.
When $\rho\simeq1$ and $\min({\rm IPR})$ starts to deviate from
zero, the system enters a localized phase and all bulk states take complex eigenvalues. 
Meanwhile, the winding number
$w_{2}$ jumps from $0$ to $-1$ whereas $w_{1}$ remains
at $-1$, as shown in Fig.~\ref{fig:PhsDiag}(a). Therefore, the transition
from the mobility edge to localized phases is also
topological and accompanied by the quantized change of winding number
$w_{2}$ by $-1$. The localized phase can then be viewed as a topological
phase with $(w_{1},w_{2})=(-1,-1)$. These observations
are all demonstrated in the phase diagram Fig.~\ref{fig:PhsDiag}(a).
We also considered other possible amplitudes of onsite potential
$V$ {[}with one example given in Fig.~\ref{fig:PhsDiag}(b){]}, and
obtain similar kinds of diagrams, which verifies the generality
of our approach to the characterization of localization transitions
and topological phases in dimerized NHQCs.
We summarize the key results of this section in Table \ref{tab:1}.
\begin{table*}
	\begin{centering}
		\begin{tabular}{|c|c|c|c|}
			\hline 
			Phase & Extended & Mobility Edge & Localized\tabularnewline
			\hline 
			\hline 
			Spectrum & Real & \multicolumn{2}{c|}{Complex}\tabularnewline
			\hline 
			Averaged AGR & $\overline{g}\simeq0$ & $0<\overline{g}<0.4$ & $\overline{g}\simeq0.4$\tabularnewline
			\hline 
			IPR & $\simeq0$ for all states & $>0$ \& $\simeq0$ coexist & $>0$ for all states\tabularnewline
			\hline 
			Winding number & $(w_{1},w_{2})=(0,0)$ & $(w_{1},w_{2})=(-1,0)$ & $(w_{1},w_{2})=(-1,-1)$\tabularnewline
			\hline 
		\end{tabular}
		\par\end{centering}
	\caption{Summary of the results for the dimerized NHQC.\label{tab:1}}
\end{table*}

\section{Discussion\label{sec:Sum}}
In experiments, the dimerized NHQC might be engineered in photonic systems.
The uniform part of hopping amplitude and non-Hermitian quasiperiodic
potential could be realized by a frequency-modulated mode-locked
laser with gain medium, phase modulator and low-finesse intracavity
etalon, as proposed in Ref.~\cite{LonghiQC1}. The dimerized hopping
amplitude could be realized by engineering the profile of refractive
index in the model-locked laser setup~\cite{ExpSetup2}. Therefore,
our model should be within reach in current or near-term experimental
situations.
To promote the detection and characterization of different
phases in the dimerized NHQC, we also investigate its wavepacket dynamics in
Appendix \ref{subsec:Dyn}, and find connections between the
dynamical signatures and localization properties of the system.

In conclusion, we find localization and topological transitions in
a dimerized NHQC, which are originated from
the cooperation between hopping dimerizations and complex onsite
quasiperiodic potential. In the region of weak dimerization and non-Hermiticity, 
the system is in an extended phase with real spectrum and
delocalized eigenstates. With the increase of hopping dimerization and
complex potential, the system transforms into a mobility edge phase.
When the strength of hopping dimerization and non-Hermitian
modulation become stronger, the system enters a third
phase in which the spectrum is complex and all eigenstates are localized.
Moreover, the transitions between the extended, mobility edge and localized
phases are of topological nature. They can be characterized
by the quantized jumps of two spectral winding numbers. 
Our results thus uncover the unique spectrum, topological
and transport features of quasicrystals due to the interplay between
hopping dimerizations and non-Hermitian onsite potential. The different
phases and transitions found in our system further reveal the
richness of localization and topological phenomena in non-Hermitian
quasicrystals. In future work, it would be interesting to
consider the impact of hopping dimerizations in other types of nonreciprocal
and non-Hermitian quasicrystals, and explore the effect of nonlinearity,
many-body interactions and skin effects on the localization and topological physics
in dimerized non-Hermitian systems.

\begin{acknowledgments}
This work is supported by the National Natural Science Foundation of China (Grant No. 11905211), the China Postdoctoral Science Foundation (Grant No. 2019M662444), the Fundamental Research Funds for the Central Universities (Grant No. 841912009), the Young Talents Project at Ocean University of China (Grant No. 861801013196), and the Applied Research Project of Postdoctoral Fellows in Qingdao (Grant No. 861905040009).
\end{acknowledgments}
\appendix
%\vspace{0.5cm}

\section{Detail of the spectrum\label{subsec:Edetail}}
\begin{figure}
	\begin{centering}
		\includegraphics[scale=0.5]{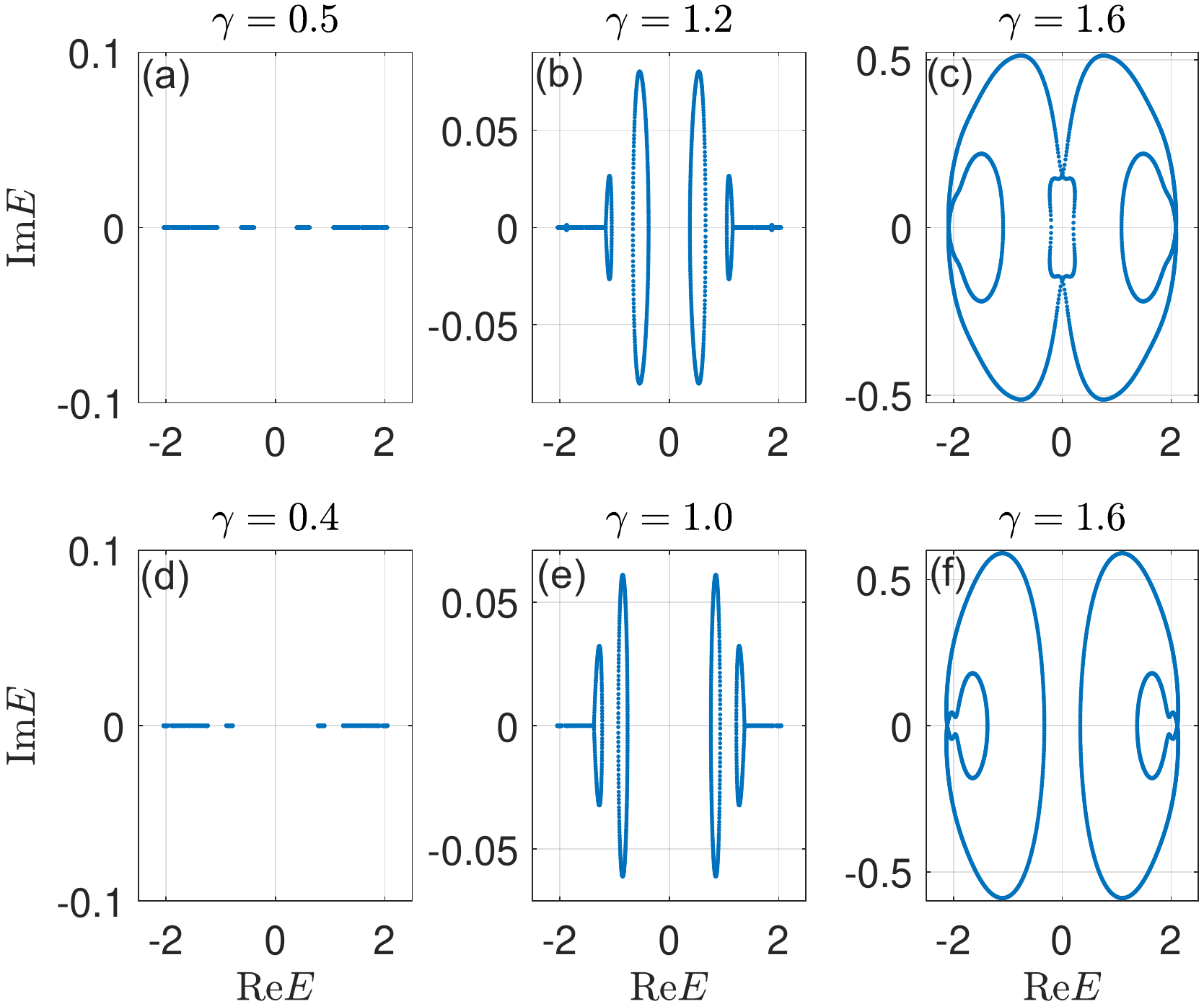}
		\par\end{centering}
	\caption{The spectrum of the dimerized NHQC model under PBC on the complex plane.
		System parameters are set as $J=1$, $V=0.5$, $\alpha=\frac{\sqrt{5}-1}{2}$,
		and $\Delta=0.2$ ($\Delta=0.4$) for (a)\textendash (c) {[}(d)\textendash (f){]}.
		The values of the imaginary part of phase shift
		are shown in the captions. The length of lattice is
		$L=2584$.\label{fig:ReEImE}}
\end{figure}

In this appendix, we provide more details about the spectrum of the dimerized NHQC.
In Fig.~\ref{fig:ReEImE},
we present the spectrum for a typical set of system parameters. 
When the imaginary part of phase shift $\gamma$ is small, we observe real
spectrum in Figs.~\ref{fig:ReEImE}(a) and \ref{fig:ReEImE}(d),
implying that the system is in the ${\cal PT}$-invariant region. With the increase
of $\gamma$, the spectrum starts to become complex and developing
loops on the complex $E$ plane, which means that the system has
undergone a ${\cal PT}$-transition and roamed into a ${\cal PT}$-broken
region. However, parts of the spectrum are still pinned to the real
axis in Figs.~\ref{fig:ReEImE}(b) and \ref{fig:ReEImE}(e),
indicating that the eigenstates with real and complex eigenvalues coexist
in these cases. When $\gamma$ further increases, the number of eigenstates
with real eigenvalues tends to decrease, and finally almost all states
have complex energies, as shown in Figs.~\ref{fig:ReEImE}(c)
and \ref{fig:ReEImE}(f). Besides, the range and shape of spectrum
also depend on the dimerization strength $\Delta$, which reveals that both the hopping
dimerization and complex onsite potential could affect the properties
of the spectrum.

\section{Lypunov exponent\label{subsec:LE}}

In this appendix, we investigate the Lypunov exponent~(LE) of the dimerized NHQC,
and compare it with the results deduced from IPRs in the main text.
The LE is the inverse of localization length, which in the thermodynamic limit
approaches zero for an extended state, and taking a finite
value for a localized state. For an eigenstate $|\psi^{(j)}\rangle=\sum_{n}\psi_{n}^{(j)}|n\rangle$
of $\hat{H}$ with energy $E_{j}$, the LE is defined in the lattice representation as
$\zeta_{j}=-\lim_{L\rightarrow\infty}\frac{1}{L}\ln|\psi_{L}^{(j)}/\psi_{1}^{(j)}|$.
Under the OBC, the $\hat{H}$
in Eq.~(\ref{eq:H}) takes a tridiagonal form in the lattice representation,
and $\zeta_{j}$ can be expressed by eigenenergies~\cite{LonghiQC2}, i.e.,
\begin{equation}
\zeta_{j}=\lim_{L\rightarrow\infty}\frac{1}{L}\left(\sum_{n=1,n\neq j}^{L}\ln|E_{j}-E_{n}|-\sum_{n=1}^{L-1}\ln|t_{n}|\right),\label{eq:LE2}
\end{equation}
where $t_{n}=J+(-1)^{n}\Delta$ is the dimerized hopping
amplitude. When the minimum of $\zeta_{j}$, i.e.,
\begin{equation}
\zeta_{\min}\equiv\lim_{L\rightarrow\infty}\left(\min_{j\in\{1,...,L\}}\zeta_{j}\right),\label{eq:LEmin}
\end{equation}
starts to deviate appreciably from zero, we expect all eigenstates
to become localized and the system enters an insulator phase. $\zeta_{\min}$
can thus be utilized to determine the onset of a fully localized phase,
just like the min(IPR) in the main text.
However, due to the existence of edge states under the OBC, one cannot
directly extract the boundary contour between extended and mobility edge
phases from the maximum of LEs, since the LE of edge states are always
nonzero even when all bulk states are extended. To resolve this issue,
we consider the LE after taking the average over all bulk and edge states, which is defined
as
$\zeta_{{\rm ave}}=\lim_{L\rightarrow\infty}\frac{1}{L}\sum_{j=1}^{L}\zeta_{j}$.
Since the number of edge states is significantly smaller then the
bulk states, we expect $\zeta_{{\rm ave}}\rightarrow0$ in the thermodynamic
limit if all bulk states are extended, and $\zeta_{{\rm ave}}>0$
when a sufficient number of bulk states ($\propto L$) become localized.
$\zeta_{{\rm ave}}$ can thus be employed to locate the boundary between
the extended and mobility edge phases. To show the boundary contour
separating these two phases more clearly, we evaluate the partial derivative
\begin{equation}
\zeta'_{{\rm ave}}\equiv\partial_{\gamma}\zeta_{{\rm ave}}.\label{eq:LEaveP}
\end{equation}
In the bulk extended phase, the number of edge states is
a constant for a given $L$, and we expect $\zeta'_{{\rm ave}}=0$.
Beyond the extended phase, localized bulk states appear and their
number increases with $\gamma$, yielding $\zeta'_{{\rm ave}}>0$.
The boundary between the regions with $\zeta'_{{\rm ave}}=0$ and $\zeta'_{{\rm ave}}>0$
should thus correspond to the boundary between extended and mobility edge
phases of the system.

\begin{figure}
	\begin{centering}
		\includegraphics[scale=0.49]{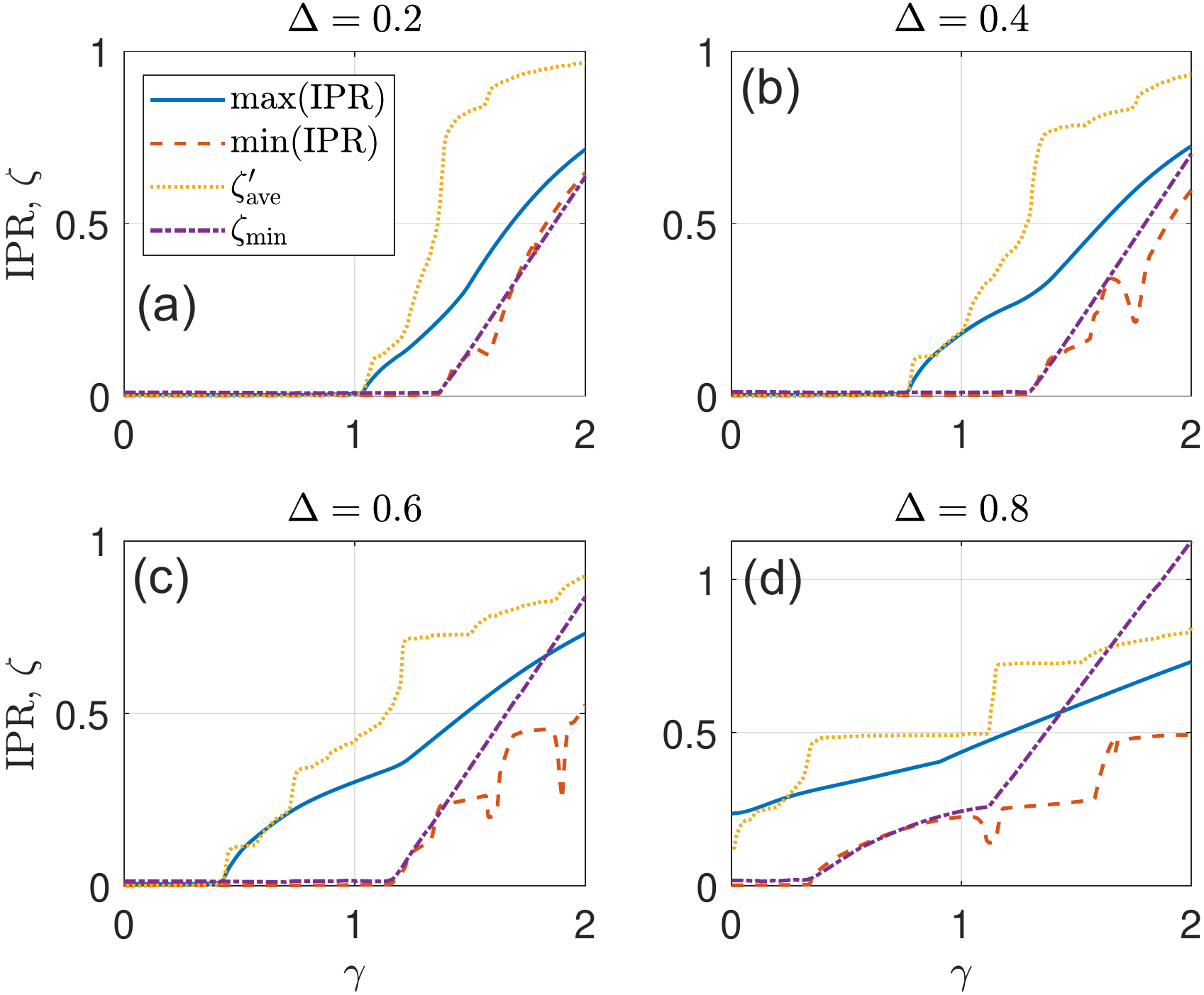}
		\par\end{centering}
	\caption{IPRs and LEs versus the imaginary part of phase shift
		$\gamma$. System parameters are $J=1$, $V=0.5$, and
		$\alpha=\frac{\sqrt{5}-1}{2}$. The length of lattice is $L=610$,
		and PBC~(OBC) is taken in the calculation of IPRs (LEs). In (a)--(d),
		the solid (dashed) lines represent the maximum (minimum) of IPRs 
		and the dotted (dash-dotted) lines show the derivative of averaged
		(minimum of) LE.\label{fig:IPRLE}}
\end{figure}

In Fig.~\ref{fig:IPRLE}, we show the IPRs and LEs versus the imaginary
part of phase shift at different hopping dimerizations. We observe that when
$\Delta$ is small, the $\max({\rm IPR})$, $\min({\rm IPR})$, $\zeta'_{{\rm ave}}$
and $\zeta_{\min}$ are all pinned to zero over a range of $\gamma$.
Compared with the results reported in Fig.~\ref{fig:DOSAGR}, it can
be verified that the spectrum of the system is real in this region.
Therefore, if $\max({\rm IPR})\simeq\min({\rm IPR})\simeq\zeta'_{{\rm ave}}\simeq\zeta_{\min}\simeq0$,
our system is in a ${\cal PT}$-invariant extended phase. 
When $\gamma$ exceeds certain critical
point $\gamma_{c1}$, both the $\max({\rm IPR})$ and $\zeta'_{{\rm ave}}$
start to deviate from zero, even though the $\min({\rm IPR})$
and $\zeta_{\min}$ are still stuck to zero. Typical forms of spectrum
of the system in this region are shown in Figs.~\ref{fig:ReEImE}(b)
and \ref{fig:ReEImE}(e). Referring to Fig.~\ref{fig:DOSAGR}, it
can be checked that the region with $\max({\rm IPR})>0,\zeta'_{{\rm ave}}>0$
and $\min({\rm IPR})\simeq\zeta_{\min}\simeq0$ also has $0<\rho<1$ and
$0<\overline{g}<0.4$. 
The system is therefore in a mobility edge phase in this region. When
$\gamma$ further increases and goes beyond a second critical point
$\gamma_{c2}$, the $\max({\rm IPR})$, $\min({\rm IPR})$, $\zeta'_{{\rm ave}}$
and $\zeta_{\min}$ all deviate from zero. According to Fig.~\ref{fig:DOSAGR},
the system has $\rho\simeq1$ and $\overline{g}\simeq0.4$ in
this region. It thus enters an insulator phase in which all bulk states
have complex energies. Note that in Figs.~\ref{fig:IPRLE}(a)\textendash (c),
the range of the mobility edge phase and the critical values $\gamma_{c1},\gamma_{c2}$
are all affected by the hopping dimerization $\Delta$. Moreover,
the extended phase could vanish when $\Delta$ is large enough,
as shown in Fig.~\ref{fig:IPRLE}(d), where we have $\max({\rm IPR})>0,\zeta'_{{\rm ave}}>0$
from the very beginning. Therefore, both the hopping dimerization
and onsite complex potential control the localization transitions
in the dimerized NHQC, and their interplay determines the complete
phase diagram of the system.

\begin{figure}
	\begin{centering}
		\includegraphics[scale=0.44]{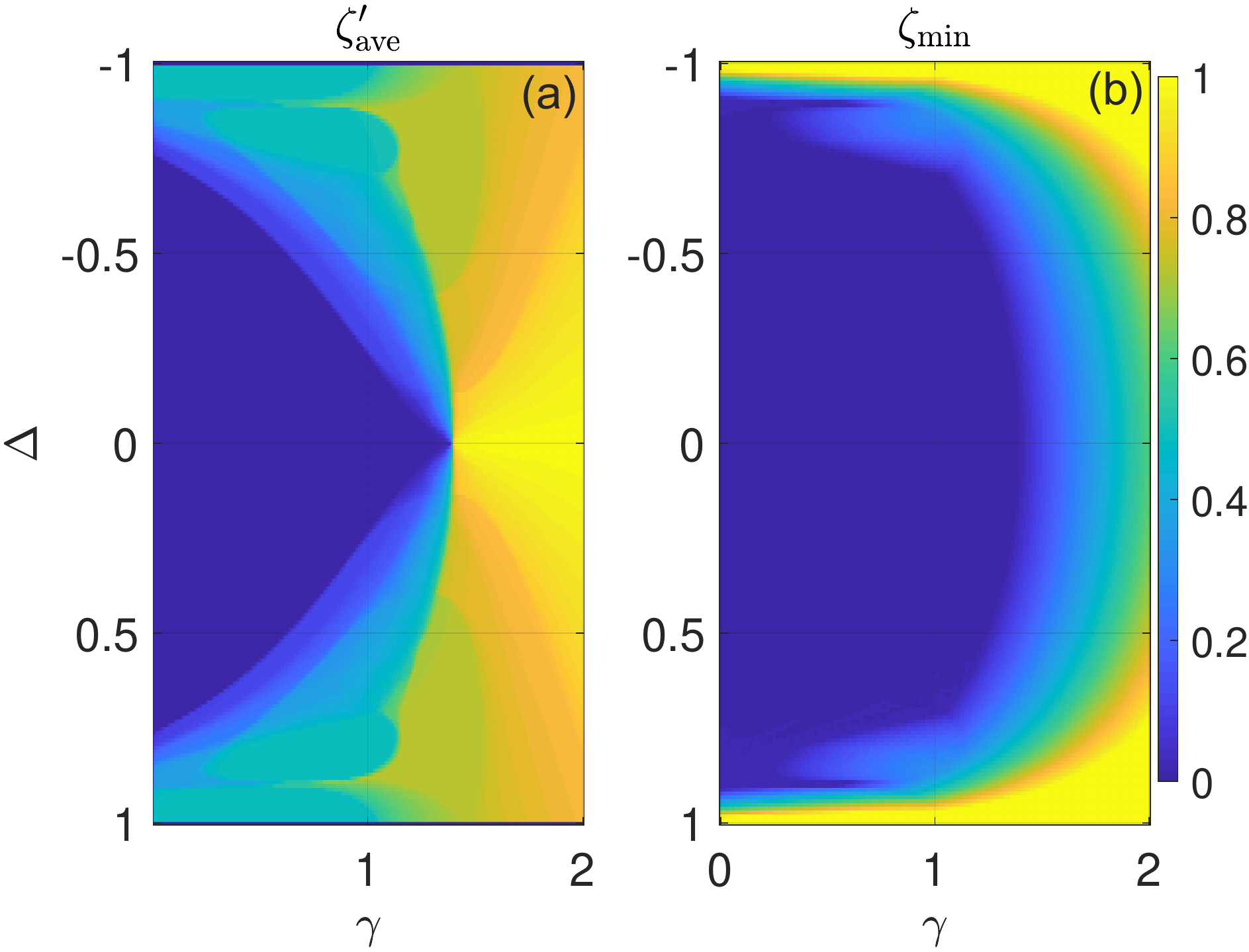}
		\par\end{centering}
	\caption{The minimum and derivative of LEs, as defined in Eqs.~(\ref{eq:LEmin})
		and (\ref{eq:LEaveP}), versus the imaginary part of phase shift $\gamma$
		and hopping dimerization $\Delta$. System parameters are set as $J=1$, $V=0.5$,
		and $\alpha=\frac{\sqrt{5}-1}{2}$. The length of lattice is $L=610$
		and the OBC is taken in the diagonalization $\hat{H}$.\label{fig:LE}}
\end{figure}

In Fig.~\ref{fig:LE}, we present the LEs
versus the imaginary part of phase shift and hopping dimerization. 
In Fig.~\ref{fig:LE}(a), we find a region with $\zeta'_{{\rm ave}}\simeq0$
(in dark blue) in the parameter space, meaning
that all states of $\hat{H}$ 
therein are extended. Notably, the scope
of this region is coincide with the region of Fig.~\ref{fig:IPR}(a)
with $\max({\rm IPR})\simeq0$, and also the regions in Fig.~\ref{fig:DOSAGR}
where the density of states with complex eigenvalues
and averaged AGR vanish. Therefore,
the region with $\zeta'_{{\rm ave}}\simeq0$ in Fig.~\ref{fig:LE}(a)
corresponds to a ${\cal PT}$-invariant extended phase, in which all eigenstates
of $\hat{H}$ have real energies. 
In Fig.~\ref{fig:LE}(b), we observe a zone
with $\zeta_{\min}\simeq0$ (in dark blue), in which extended
states of $\hat{H}$ persist up to its boundary. 
The zone with $\zeta_{\min}\simeq0$
covers the region with $\zeta'_{{\rm ave}}\simeq0$.
Beyond that, we find a region with $\zeta_{\min}\simeq0$
and $\zeta'_{{\rm ave}}>0$ before the system entering a localized
phase. The scope of
this region is consistent with the region in Fig.~\ref{fig:IPR}
with $\min({\rm IPR})\simeq0$, $\max({\rm IPR})>0$, and
with the region in Fig.~\ref{fig:DOSAGR} with $0<\rho<1$, $0<\overline{g}<0.4$,
verifying that there indeed exists a mobility edge phase
in the dimerized NHQC.

\section{Wavepacket dynamics\label{subsec:Dyn}}

In this appendix, we investigate
the dynamical properties of the dimerized NHQC,
and suggest to characterize the different phases
by the spreading velocity and return probability of wavepackets. 
The evolution of a state in our system
is obtained by solving the time-dependent Schr\"odinger equation, yielding 
$|\tilde{\psi}(t)\rangle=e^{-i\hat{H}t}|\psi(0)\rangle$.
Since $\hat{H}$ is non-Hermitian, the evolution can be nonunitary
and the state $|\tilde{\psi}(t)\rangle$ is not normalized. The normalized
state is given by 
$|\psi(t)\rangle=|\tilde{\psi}(t)\rangle/\sqrt{\langle\tilde{\psi}(t)|\tilde{\psi}(t)\rangle}$,
which can be used to study the probability distribution and expectation
values of observables. To characterize
the spreading of a wavepacket, we consider its second
moment, which in the lattice representation reads
$\sigma_{2}=\sum_{n=1}^{L}n^{2}|\psi_{n}(t)|^{2}$.
Here $n$ is the lattice index and $\psi_{n}(t)=\langle n|\psi(t)\rangle$.
From the second moment, we can further obtain the spreading velocity
of the wavepacket by averaging over a time duration~\cite{LonghiQC4},
i.e., $v(t)=\sqrt{\sigma_{2}}/t$,
which could behave differently if the wavepacket is initialized in
different phases of the system. In the extended phase, 
an initially localized wavepacket is expected
to undergo ballistic spreading, forming a light cone pattern on the
space-time plane if the hopping is symmetric~\cite{LonghiQC4}. The
velocity $v(t)$ of the wavepacket would be finite.
In an insulator phase, an initially
localized wavepacket will refuse to spread and stay around
its original site. Its spreading velocity $v(t)$ will also vanish,
showing the phenomena of dynamical localization~\cite{LonghiQC4}.

\begin{figure}
	\begin{centering}
		\includegraphics[scale=0.46]{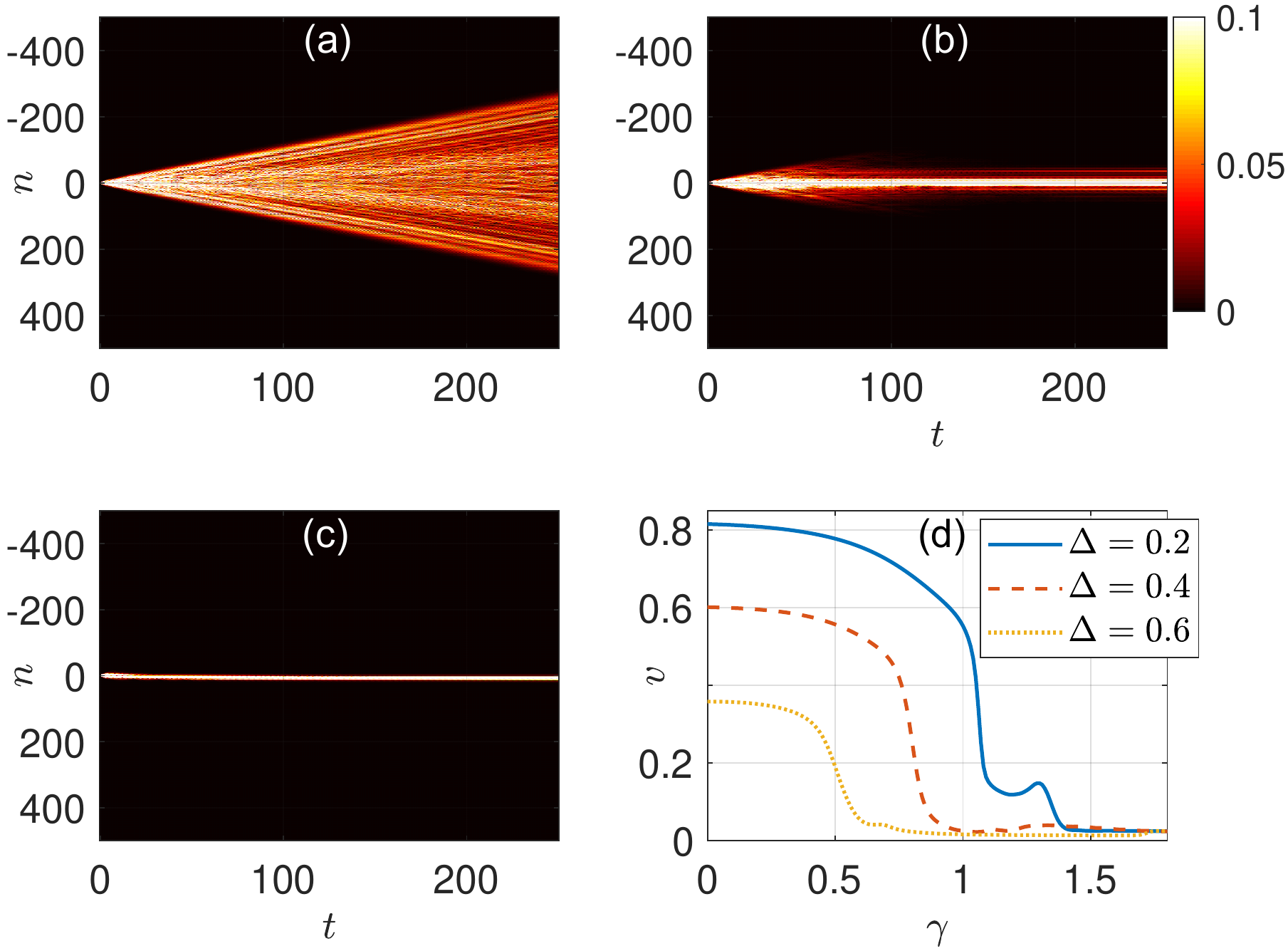}
		\par\end{centering}
	\caption{Profiles (a)\textendash (c) and spreading
		velocities (d) of wavepackets initially localized at
		the central site of the lattice. The imaginary phase shift is set
		as (a) $\gamma=0.4$, (b) $\gamma=1$ and (c) $\gamma=1.6$.
		System parameters are $J=1$, $V=0.5$, $\Delta=0.4$ in
		(a)\textendash (c). The length of lattice is $L=1000$ with PBC.
		The spreading velocity $v$ is obtained
		in (d) after averaging $v(t)$ over a time span of
		$t=250$.\label{fig:PsiVT}}
\end{figure}

In Fig.~\ref{fig:PsiVT}, we show the profiles of wavepacket
during the evolution and the averaged spreading velocity
for a typical set of system parameters.
In Fig.~\ref{fig:PsiVT}(a), the system is set in the
extended phase (see also Fig.~\ref{fig:WN}).
We observe that the wavepacket indeed performs
a ballistic spreading and presents a light cone pattern. In Fig.~\ref{fig:PsiVT}(b),
the system is set in the mobility edge phase (see also Fig.~\ref{fig:WN}).
The wavepacket is initially found to undergo
a slow spreading over a finite range of sites. At a later
stage, the spreading tends to terminate and the wavepacket retains
a finite width around its original site. The initial spreading
process may be assisted by the remaining extended states of the mobility edge
phase. But during the evolution, the wavepacket develops
more overlap with the localized states, which finally help to
shut off its transport in the lattice. In Fig.~\ref{fig:PsiVT}(c),
the parameters of $\hat{H}$ are set in the localized phase (see also
Fig.~\ref{fig:WN}). As expected,
the wavepacket could not spread and remains exponentially localized
around its original location. The distinctive signatures of wavepacket
spreading in these three phases could then provide us with a dynamical
way of distinguishing them. In Fig.~\ref{fig:PsiVT}(d),
the spreading velocity of a wavepacket is found to decrease with the
increase of $\gamma$ when the system is set in the extended phase.
At large $\gamma$, the system enters the localized phase and the
velocity approaches zero. When $\gamma$ goes across the boundary
between the extended and mobility edge phases, the velocity
shows a quick drop in a small range of $\gamma$, implying the
appearance of a dynamical transition between these two phases. Notably,
the critical point $\gamma_{c1}$ of the transition is different for
different hopping dimerizations, and the velocity
could also show an anomalous growth with the increase of $\gamma$
in the mobility edge phase {[}e.g., see the solid line in Fig.~\ref{fig:PsiVT}(d){]}.
These observations suggest that the interplay between hopping
dimerizations and complex onsite potential could not only create new
types of dynamical phases in non-Hermitian quasicrystals, but also
cause non-Hermiticity enhanced transport observed previously in
systems with nonreciprocal hopping~\cite{LonghiQC4}.

\begin{figure}
	\begin{centering}
		\includegraphics[scale=0.46]{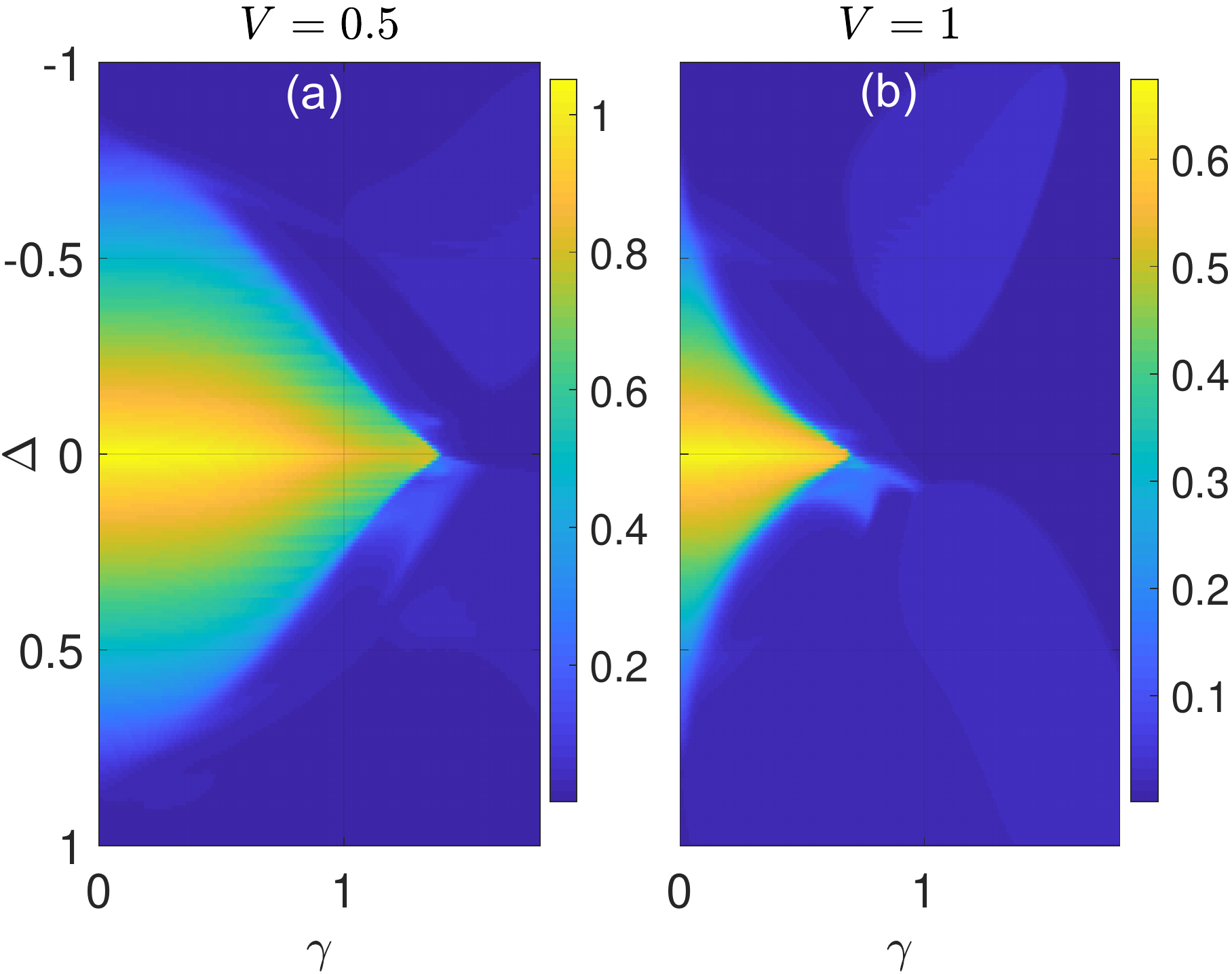}
		\par\end{centering}
	\caption{Spreading velocities of an initially localized wavepacket versus
		the imaginary part of phase shift $\gamma$ and hopping dimerization
		$\Delta$. $v(t)$ is averaged over a time duration
		$t=250$. The strength of the onsite potential is $V=0.5$ in (a)
		and $V=1$ in (b). System parameters are set as $J=1$
		and $\alpha=\frac{\sqrt{5}-1}{2}$. The length of lattice is $L=1000$ with PBC,
		and the initial state satisfies $\langle n|\psi(0)\rangle=\delta_{n0}$. 
		\label{fig:VT}}
\end{figure}

In Fig.~\ref{fig:VT}, we plot the mean velocities of
initially localized wavepackets versus the imaginary part of phase
shift and hopping dimerization. In Figs.~\ref{fig:VT}(a)
and \ref{fig:VT}(b), we find clear borders between
regions in which $v$ take finite values and approaches zero.
They are coincide with the boundaries between extended
and mobility edge phases, as shown in Figs.~\ref{fig:PhsDiag}(a)
and \ref{fig:PhsDiag}(b). Therefore, the
velocity of wavepackets could be employed to locate the boundaries
between extended and mobility edge phases. However,
we could not observe a clear dynamical signature of the boundary between
mobility edge and localized phases, as they both contain a sufficient
amount of localized states to block the transport. One
way to distinguish these two phases is by investigating the detailed
profile of wavepackets during the evolution, as shown in Figs.~\ref{fig:PsiVT}(a)\textendash (c).
Another way is to consider
the return probability of a state following quantum quenches~\cite{ChenQC5},
defined as
$P(t)=|\langle\psi^{{\rm i}}|\psi^{{\rm f}}(t)\rangle|^{2}$,
where the initial state $|\psi^{{\rm i}}\rangle$ is an eigenstate
of the prequench Hamiltonian $\hat{H}^{{\rm i}}$, and
$|\psi^{{\rm f}}(t)\rangle$ is the final state
evolved by the postquench Hamiltonian $\hat{H}^{{\rm f}}$
(not commute with $\hat{H}^{{\rm i}}$) from $t=0$ to $t$,
i.e.,
$|\psi^{{\rm f}}(t)\rangle=|\tilde{\psi}^{{\rm f}}(t)\rangle\sqrt{\langle\tilde{\psi}^{{\rm f}}(t)|\tilde{\psi}^{{\rm f}}(t)\rangle}$
and $|\tilde{\psi}^{{\rm f}}(t)\rangle=e^{-i\hat{H}^{{\rm f}}t}|\psi^{\rm i}\rangle$.
For a given initial state, if the parameters of the pre and postquench
Hamiltonians are set in phases with different localization
nature, we expect the $P(t)$ to show qualitatively different
behaviors. One can thus distinguish any two different phases of the
system by investigating the wavepacket dynamics following a quench
across the boundary between them. In our case, we choose the initial state
to be an eigenstate of ${\hat H}$ with $\gamma=\gamma_{\rm i}$, and perform
a quench at $t=0$ by letting $\gamma_{\rm i}\rightarrow\gamma_{\rm f}$.

\begin{figure}
	\begin{centering}
		\includegraphics[scale=0.48]{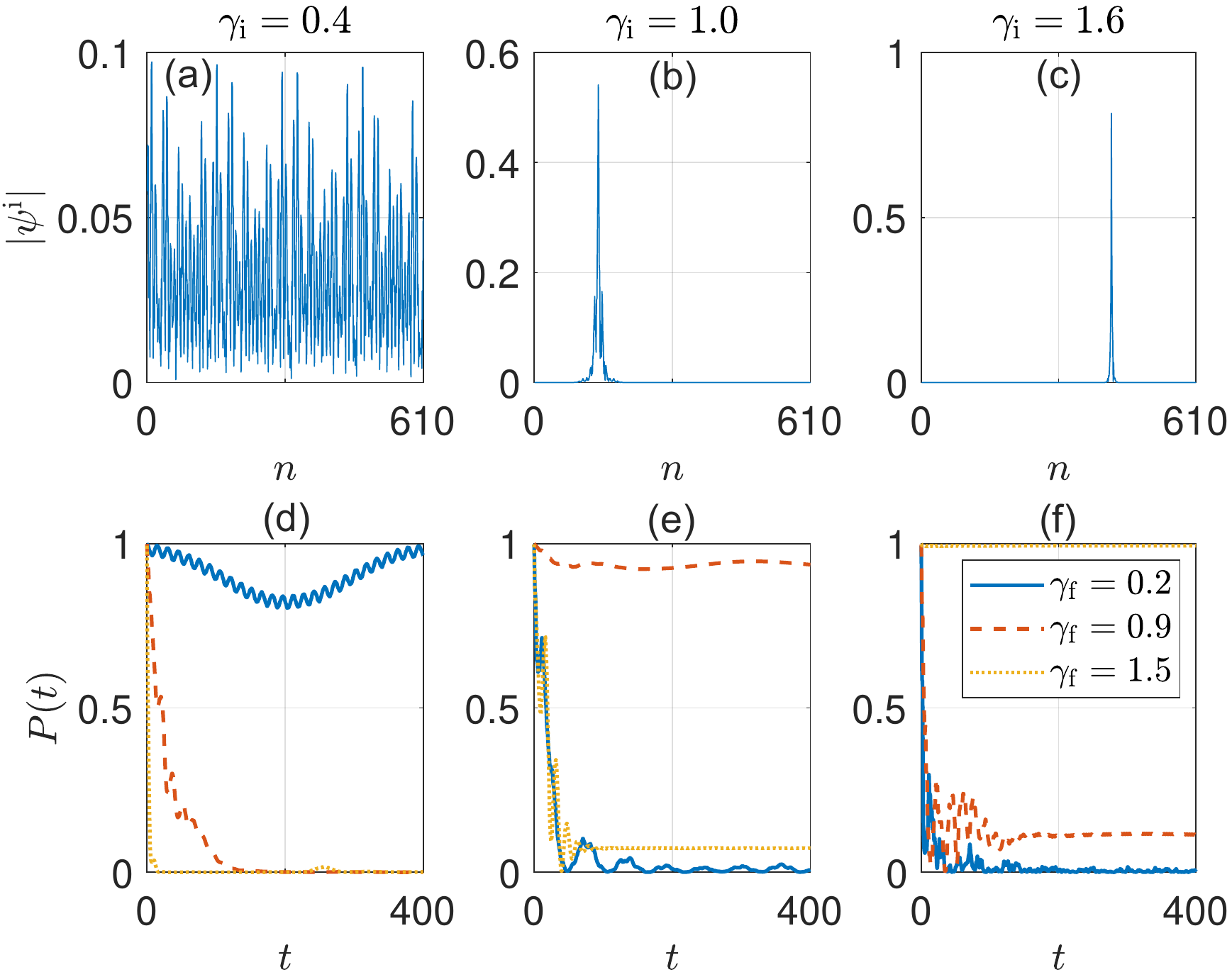}
		\par\end{centering}
	\caption{Profiles of initial states {[}in (a)\textendash (c){]} and
		their return probabilities following quenches {[}in (d)\textendash (f){]}
		in the dimerized NHQC. The states in (a)\textendash (c)
		are chosen to be the ones with the largest imaginary parts of energy
		at the corresponding phase shift $\gamma_{{\rm i}}$.
		System parameters for (a)\textendash (c) are set as $J=1$,
		$V=0.5$, $\Delta=0.4$, $\alpha=\frac{\sqrt{5}-1}{2}$, and the length
		of lattice is $L=610$ with PBC. In (d)--(f), the curves correspond
		to return probabilities of the initial states in (a)--(c)
		following quenches of $\gamma$ from the corresponding $\gamma_{{\rm i}}$ to different
		values of $\gamma_{{\rm f}}$ applied at $t=0$.\label{fig:RP}}
\end{figure}

In Fig.~\ref{fig:RP}, we show the return probabilities of initial
states prepared in and quenched to different phases of the dimerized
NHQC. In Fig.~\ref{fig:RP}(a), the system is initialized in
the extended phase with state profile
$|\psi^{{\rm i}}|$. When the postquench system is in
the same phase, $P(t)$ 
undergoes oscillations without a global decay, as shown
by the blue solid line in Fig.~\ref{fig:RP}(d). If the postquench
Hamiltonian is in the mobility edge phase,
$P(t)$ first subject to oscillations with a global
decay profile, and approaches zero in the long-time
domain, as demonstrated by the red dashed line in Fig.~\ref{fig:RP}(d).
The initial oscillations of $P(t)$ can be traced back
to the presence of extended states in the mobility edge phase. With
the progress of time, the state is trapped by localized
states in the mobility edge phase, leading to the decay of $P(t)$
at large $t$. If $\hat{H}^{{\rm f}}$ is in the localized
phase, $P(t)$ decays very
fast and quickly approaches zero, as depicted by the
yellow dotted line in Fig.~\ref{fig:RP}(d). Since all states are
localized in the postquench system, the evolving state can never goes
back. The distinctive features of
$P(t)$ in three phases can thus help us to discriminate them
following the postquench dynamics if the initial state is prepared
in the extended phase.

In Figs.~\ref{fig:RP}(b) and \ref{fig:RP}(c), we prepare the system 
into a localized state in the mobility edge and insulator phases.
The results in Figs.~\ref{fig:RP}(e) and \ref{fig:RP}(f)
suggest that the return probabilities also behave distinctly
when $\hat{H}^{\rm f}$ is set in phases with different transport
nature. Putting together, we conclude that
wherever the initial state is prepared, its return probability
following quenches to different phases could
help us to distinguish them dynamically. In experiments, the
averaged spreading velocity and return probability of wavepackets
can thus be employed to detect phases with different localization
nature in the dimerized NHQC.

%\color{black}
%\newpage{}

\end{document}